# Calculation of Residual Dose around Small Objects Using Mu2e Target as an Example[*]

V.S. Pronskikh[#], A.F. Leveling, N.V. Mokhov, I.L. Rakhno

Fermi National Accelerator Laboratory, Batavia, IL 60510, USA

P. Aarnio

Department of Applied Physics, Aalto University, FI-00076, Finland

## Abstract

The MARS15 code provides contact residual dose rates for relatively large accelerator and experimental components for predefined irradiation and cooling times. The dose rate at particular distances from the components, some of which can be rather small in size, is calculated in a post Monte-Carlo stage via special algorithms described elsewhere. The approach is further developed and described in this paper.

---

[*]Work supported by Fermi Research Alliance, LLC under contract No. DE-AC02-07CH11359 with the U.S. Department of Energy.
[#]vspron@fnal.gov

# Calculation of Residual Dose around Small Objects Using Mu2e Target as an Example

V.S. Pronskikh, A.F. Leveling, N.V. Mokhov, I.L. Rakhno
Fermi National Accelerator Laboratory, Batavia, IL 60510, USA

P. Aarnio
Department of Applied Physics, Aalto University, FI-00076, Finland

## INTRODUCTION

The MARS15 code [1] provides contact residual dose rates for relatively large accelerator and experimental components for predefined irradiation and cooling times. Relatively large means that a characteristic dimension of the object is much larger than the mean free path of ~1 MeV photons emitted from the nuclides produced in the object [2]. For practical applications, one needs to know the dose rate at particular distances from the components, some of which can be rather small in size. This is done in a post Monte-Carlo stage via algorithms developed in Ref. [2, 3]. In this paper the approach is further developed.

Two methods are described and compared using as an example the Mu2e target station. The first method implies (1) MARS15 calculation of the residual dose on contact with the target, (2) correction for a small target size, and (3) distance correction which is also Monte-Carlo based. The second method is based on (1) MARS15 calculation of the production rates of individual residual nuclei produced both in spallation reactions in the target and in consequent decay chains, (2) calculation of activities of the isotopes in the target with the DeTra code [4] after certain irradiation and cooling times, and (3) conversion of activities to individual doses at a distance using specific gamma-ray constants, some of which were calculated in this work.

## FIRST METHOD

In the Mu2e experiment, the 8-GeV proton beam would hit a tilted gold target at a rate of 2.e13 p/s. (Note that this intensity will be probably 6 times lower at the start of the experiment). The target (16 cm long and 0.6 cm in diameter) is surrounded by a 0.03 cm layer of cooling water and 0.05 cm of Ti. Residual dose on contact is calculated using MARS15 with all residual nuclei generated in the target effectively taken into account. Average dose after one year of irradiation and one week of cooling is found to be 20 kSv/hr (see Fig. 1).

*Correction for small target radius.* As described in Ref. [2], one needs to apply a dose correction factor if the target is relatively small. A graph to determine such a correction is given in Fig. 2. It was calculated in Ref. [2] using Monte-Carlo as a function of the parameter $x_t$ – ratio of the target diameter to the mean free path of the 1 MeV photons ($\lambda_t$ (Au)≈ 0.745 cm). For the Mu2e target considered here, this dose factor is $R_G$=0.41.

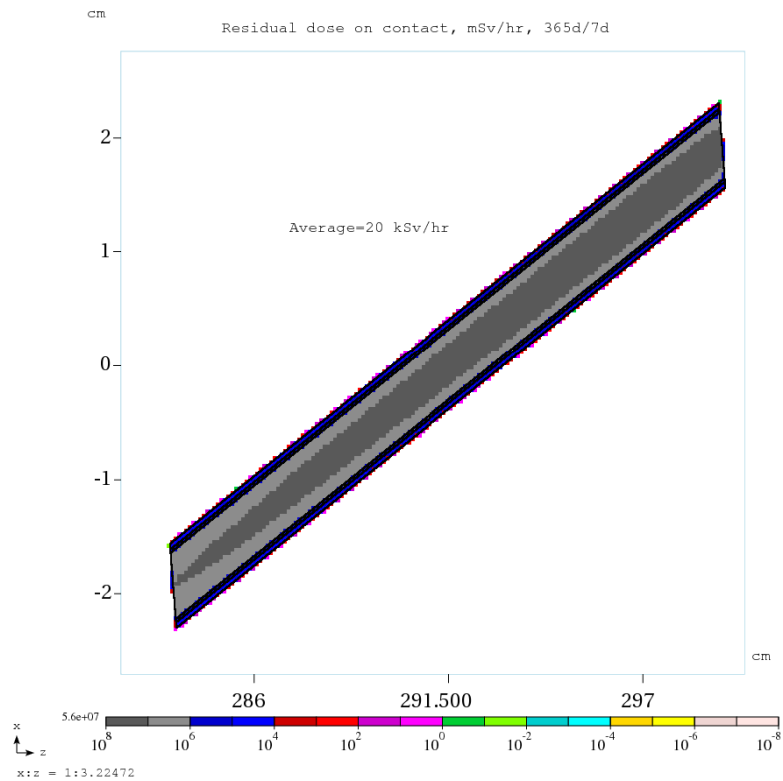

Figure 1. Residual dose (mSv/hr) on contact in the target after a year of irradiation and a week of cooling.

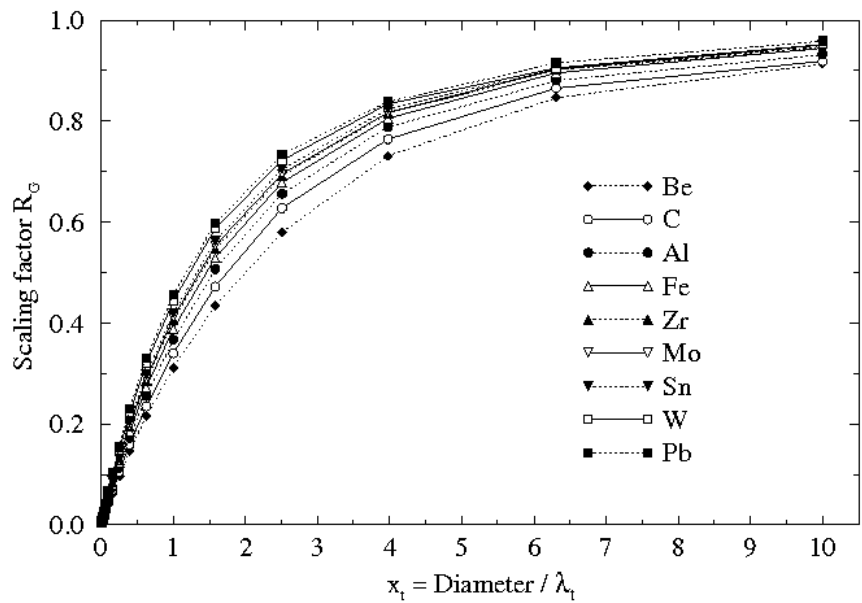

Figure 2. Dose scaling factors for solid cylinders of various materials as a function of normalized diameter (Fig. 5 from Ref. [2]).

***Dose at a distance from the target.*** The residual dose D(x,y,z) in the air due to residual activity of an irradiated object assuming isotropic angular distribution of γ-rays emitted from the surface of the object, can be described with the following equation:

$$D(x,y,z) = k_d \emptyset(x,y,z) = k_d \int dS \frac{A_s}{2\pi\rho^2}, \quad (1)$$

where $\emptyset(x,y,z)$ is flux of γ-rays, $A_s$ is surface emission rate of γ-rays per unit area and per $2\pi$ solid angle, $\rho$ is distance between the observation point and the surface element $dS$, and $k_d$ is flux-to-dose conversion factor. Typical photon energy is ~1 MeV [2]; therefore for simplicity Equation (1) does not take into account the energy dependence.

If we assume [3] the target to be an uniformly activated infinite cylindrical object, make use of the symmetry of the problem, and write the integral in the closed form, dose D(r) at a distance r from the infinite cylindrical target can be described by the equation: $D(r) = \frac{D_0}{1+\frac{r}{R}} F(\varphi\backslash\alpha)$, where $D_0$ – dose on contact with the target, $r = \sqrt{x^2+y^2+z^2}$, R – radius of the cylinder, $F(\varphi\backslash\alpha) = \int_0^\varphi (1 - sin^2\alpha\, sin^2\theta)^{-\frac{1}{2}} d\theta$, incomplete elliptic integral of the first kind, $\varphi = \sin^{-1}\sqrt{\frac{1+R/r}{2}}$, $\alpha = \sin^{-1}\left(\frac{2\sqrt{r/R}}{1+r/R}\right)$.

Determined this way, the dose attenuation factors $f(r-R) \equiv \frac{D(r)}{D_0}$ for the lateral distances 30 and 100 cm from the infinitely long cylinder of the small radius considered here are f(30) = 0.0078 and f(100) = 0.0023, respectively.

***Correction for finite target length.*** The procedure described above is applied to an infinitely long cylindrical target, whereas the actual Mu2e target has the length of only 16 cm. The necessary correction is based on the gamma flux determination at the distance of the expected detector with the actual (16 cm) and quasi-infinite (20 m) targets and using the ratio of these fluxes as the correction factor. Another correction factor which was necessary to apply in this case is that for the different target volumes (the actual and the quasi-infinite ones), which is 125 for our model. Photon flux and dose were calculated with MARS15 for these two targets in a 1-cm wide and 1-cm thick tissue-equivalent bands positioned around the target centers at 30 and 100 cm from the target axis. The photon flux isocontours are shown in Fig. 3.

The finite size correction factors were found to be $f_s = \frac{\emptyset^{short}}{\emptyset^{long}*k_{vol}}$, $f_s(30)$=0.203, $f_s(100)$=0.0611. Starting from the average dose on contact of 20 kSv/hr, ***the resulting dose*** rates employing all the scaling factors derived above are as follows

Dose at r=30 cm:
$$D(30)=20000*0.41*0.0078*0.203=12.98 \text{ Sv/hr.}$$

Dose at r=100 cm:
$$D(100)=20000*0.41*0.0023*0.0611 = 1.15 \text{ Sv/hr.}$$

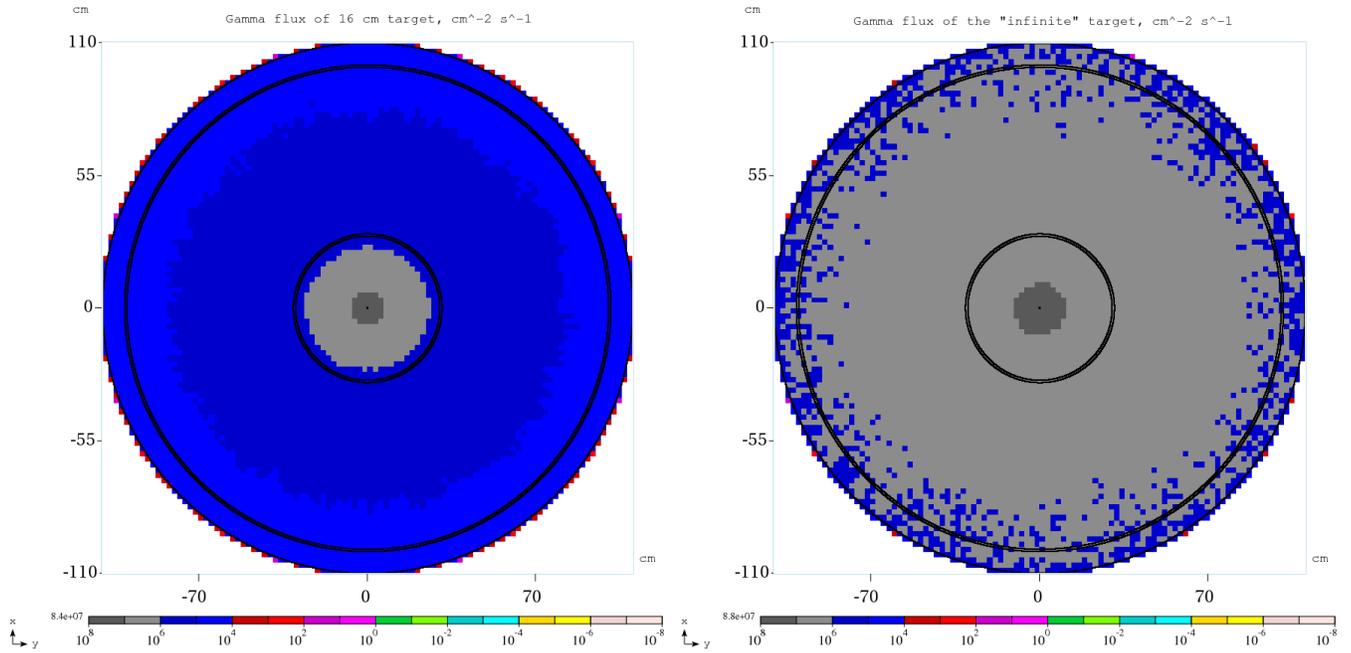

Figure 3. Photon flux around short (left) and quasi-infinite (right) targets, corrected for volume.

## SECOND METHOD

Production rates of the residual nuclides in the gold target are again calculated using MARS15. These production rates are used as initial data for the DeTra code for one year of irradiation and one week of cooling.

DeTra [4] solves the Bateman equations governing the decay and transmutation of nuclides using transmutation trajectory analysis (TTA). The core of the method is that a complex web of decay and transmutation reactions can be decomposed into a set of linear chains consisting of all possible routes, or trajectories, through the web. A set of linear chains is constructed for each nuclide and following all the possible reaction and decay modes leading to it. The concentrations of nuclides encountered in each chain are calculated by assuming that only the first nuclide of the chain has non-zero initial atomic density. Doing this for each nuclide in the initial composition and superposing the results yields the solution of the original problem. TTA is an analytic solution and thus, in principle, accurate, i.e., within the accuracy of the numerical solution. However, some chains may become very long and are thus cut when the contribution falls below a given threshold. In the case considered here, the chains are not very long and not cyclic.

From the residual isotopes we selected 19 with the highest activities that contribute >1% to the total activity (cumulatively 70%). Activities were converted to dose at r=30 cm (1 foot) using specific gamma-ray constants, $\Gamma$ [rem/(hr-MBq)], some of which were taken from [6], and the others were calculated using the empirical rule: $\Gamma = 6 * E_\gamma * I_\gamma$, where $E_\gamma$ – gamma-ray energy (MeV), and $I_\gamma$ – gamma-ray activity per decay (Ci). Doses at one meter were obtained from doses at one foot calculated using this method employing the inverse squares rule. Table 1 shows doses for individual isotopes.

Table 1. Doses calculated for individual residual isotopes based on their specific gamma ray constants.

| Isotope | Activity, Bq | Γ, mSv/(h-MBq) @ 1m | Dose at 1 meter, mSv | Dose at 1 foot, mSv | Source of the specific gamma-ray constant, Γ |
|---|---|---|---|---|---|
| Eu-146 | 3.47E+10 | 2.52E-04 | 8.75E+00 | 1.03E+02 | calculated |
| Yb-169 | 3.10E+11 | 8.84E-05 | 2.74E+01 | 2.95E+02 | [6] |
| Lu-171 | 2.14E+11 | 8.71E-05 | 1.86E+01 | 2.00E+02 | calculated |
| Lu-172 | 1.24E+11 | 2.86E-04 | 3.56E+01 | 3.83E+02 | calculated |
| Hf-175 | 4.91E+11 | 4.52E-05 | 2.22E+01 | 2.39E+02 | calculated |
| Hf-178m | 1.89E+11 | 1.46E-04 | 2.75E+01 | 2.96E+02 | calculated |
| Ta-178 | 5.50E+11 | 8.83E-06 | 4.86E+00 | 5.23E+01 | calculated |
| Hf-179m | 1.39E+11 | 3.10E-05 | 4.31E+00 | 1.74E-22 | calculated |
| Re-184 | 1.03E+11 | 1.57E-04 | 1.62E+01 | 1.74E+02 | [6] |
| Os-185 | 1.10E+12 | 1.31E-04 | 1.44E+02 | 1.55E+03 | [6] |
| Ir-188 | 8.84E+11 | 4.66E-04 | 4.12E+02 | 4.43E+03 | calculated |
| Pt-188 | 7.21E+11 | 2.04E-05 | 1.47E+01 | 1.58E+02 | calculated |
| Ir-189 | 1.36E+12 | 3.77E-06 | 5.11E+00 | 5.50E+01 | calculated |
| Ir-190 | 2.96E+11 | 2.68E-04 | 7.95E+01 | 8.55E+02 | [6] |
| Pt-191 | 4.11E+11 | 6.57E-05 | 2.70E+01 | 2.91E+02 | ORNL |
| Ir-192 | 3.87E+11 | 1.60E-04 | 6.19E+01 | 6.66E+02 | [6] |
| Au-194 | 6.43E+10 | 1.78E-04 | 1.15E+01 | 1.24E+02 | [6] |
| Au-195 | 1.61E+12 | 2.36E-05 | 3.80E+01 | 4.09E+02 | [6] |
| Au-196 | 1.31E+12 | 9.92E-05 | 1.30E+02 | 1.40E+03 | [6] |
| | | Sum dose | 1.09E+03 | 1.17E+04 | |

Using this method the dose rates were found to be 11.7 Sv/hr at one foot and 1.09 Sv/hr at one meter. After the correction for the total activity, the total dose at 30 cm from the target became D(30) = 16.7 Sv/hr, and at 100 cm from the target it became D(100) = 1.56 Sv/hr. These results are about 30% higher than the ones obtained by the first method. Doses calculated by both approaches for the Mu2e target are summarized in Table 2. Note that these values will be six times lower for the Mu2e running at one sixth of the original intensity of 2.e13 p/s.

Table 2. Summary table of doses calculated using the two methods.

| Distance from the target, cm | Dose, Sv/hr (first method) | Dose, Sv/hr (second method) |
|---|---|---|
| 30 | 12.98 | 16.7 |
| 100 | 1.15 | 1.56 |

# CONCLUSION

Two methods to calculate the residual dose rates around small targets starting from the contact dose or nuclide production rate generated by MARS15 are described. The first method employs MARS15 for the calculation of the residual dose on contact, and then uses scaling factors to correct for the target size, incomplete elliptic integral of the first kind to introduce the distance correction and the Monte-Carlo-based finite size correction. The second one uses production rates for residual isotopes calculated by MARS15 as an input for DeTra, then activities are converted to dose at one foot using specific gamma-ray constants. Both methods reveal a good agreement. These dose values can be used to predict personnel radiation dose rates for activities to be conducted in the proposed facility.